
\parskip=\medskipamount
\overfullrule=0pt
\raggedbottom
\def\normalparindent{24pt}
\nopagenumbers
\footline={\ifnum\pageno=1{\hfil}\else{\hfil\rm\folio\hfil}\fi}
\def\endpage{\vfill\eject}
\def\beginlinemode{\endmode\begingroup\parskip=0pt
                   \obeylines\def\\{\par}\def\endmode{\par\endgroup}}
\def\beginparmode{\endmode\begingroup \def\endmode{\par\endgroup}}
\let\endmode=\par
\def\raggedcenter{
                  \leftskip=2em plus 6em \rightskip=\leftskip
                  \parindent=0pt \parfillskip=0pt \spaceskip=.3333em
                  \xspaceskip=.5em\pretolerance=9999 \tolerance=9999
                  \hyphenpenalty=9999 \exhyphenpenalty=9999 }
\def\oneandahalfspace{\baselineskip=\normalbaselineskip
  \multiply\baselineskip by 3 \divide\baselineskip by 2}
\def\\{\cr}
\let\rawfootnote=\footnote\def\footnote#1#2{{\parindent=0pt\parskip=0pt
        \rawfootnote{#1}{#2\hfill\vrule height 0pt depth 6pt width 0pt}}}
\def\title{\null\vskip 3pt plus 0.2fill\beginlinemode\raggedcenter\bf}
\def\author{\vskip 3pt plus 0.2fill \beginlinemode\raggedcenter}
\def\affil{\vskip 3pt plus 0.1fill\beginlinemode\raggedcenter\it}
\def\abstract{\vskip 3pt plus 0.3fill \beginparmode{\noindent  ABSTRACT:~}  }
\def\endtitlepage{\endpage\body}
\def\body{\beginparmode\parindent=\normalparindent}
\def\head#1{\par\goodbreak{\immediate\write16{#1}
           {\noindent\bf #1}\par}\nobreak\nobreak}

\def\refto#1{$^{[#1]}$}
\def\ref#1{Ref.~#1}
\def\Ref#1{Ref.~#1}\def\cite#1{{#1}}\def\[#1]{[\cite{#1}]}

\def\(#1){(\call{#1})}
\def\call#1{{#1}}\def\taghead#1{{#1}}
\def\references{\head{REFERENCES}\beginparmode\frenchspacing\parskip=0pt}
\gdef\refis#1{\item{#1.\ }}
\def\endreferences{\body}
\def\endit{\endmode\vfill\supereject}\let\endpaper=\endit
\def\gsim{\mathrel{\raise.3ex\hbox{$>$\kern-.75em\lower1ex\hbox{$\sim$}}}}
\def\lsim{\mathrel{\raise.3ex\hbox{$<$\kern-.75em\lower1ex\hbox{$\sim$}}}}
\def\sla{\raise.15ex\hbox{$/$}\kern-.72em}
\def\iafedir{Instituto de Astronom\'\i a y F\'\i sica del Espacio\\Casilla de
          Correo 67 - Sucursal 28, 1428 Buenos Aires, Argentina}
\def\fceyn{Departamento de F{\'\i}sica\\
      Facultad de Ciencias Exactas y Naturales, Universidad de Buenos Aires\\
      Ciudad Universitaria - Pabell\'on I, 1428 Buenos Aires, Argentina}
\catcode`@=11
\newcount\r@fcount \r@fcount=0\newcount\r@fcurr
\immediate\newwrite\reffile\newif\ifr@ffile\r@ffilefalse
\def\w@rnwrite#1{\ifr@ffile\immediate\write\reffile{#1}\fi\message{#1}}
\def\writer@f#1>>{}
\def\referencefile{\r@ffiletrue\immediate\openout\reffile=\jobname.ref%
  \def\writer@f##1>>{\ifr@ffile\immediate\write\reffile%
    {\noexpand\refis{##1} = \csname r@fnum##1\endcsname = %
     \expandafter\expandafter\expandafter\strip@t\expandafter%
     \meaning\csname r@ftext\csname r@fnum##1\endcsname\endcsname}\fi}%
  \def\strip@t##1>>{}}

\def\citeall#1{\xdef#1##1{#1{\noexpand\cite{##1}}}}
\def\cite#1{\each@rg\citer@nge{#1}}
\def\each@rg#1#2{{\let\thecsname=#1\expandafter\first@rg#2,\end,}}
\def\first@rg#1,{\thecsname{#1}\apply@rg}
\def\apply@rg#1,{\ifx\end#1\let\next=\relax%
\else,\thecsname{#1}\let\next=\apply@rg\fi\next}%
\def\citer@nge#1{\citedor@nge#1-\end-}
\def\citer@ngeat#1\end-{#1}
\def\citedor@nge#1-#2-{\ifx\end#2\r@featspace#1
  \else\citel@@p{#1}{#2}\citer@ngeat\fi}
\def\citel@@p#1#2{\ifnum#1>#2{\errmessage{Reference range #1-#2\space is bad.}
    \errhelp{If you cite a series of references by the notation M-N, then M and
    N must be integers, and N must be greater than or equal to M.}}\else%
{\count0=#1\count1=#2\advance\count1 by1\relax\expandafter\r@fcite\the\count0,%
  \loop\advance\count0 by1\relax
    \ifnum\count0<\count1,\expandafter\r@fcite\the\count0,%
  \repeat}\fi}
\def\r@featspace#1#2 {\r@fcite#1#2,}    \def\r@fcite#1,{\ifuncit@d{#1}
    \expandafter\gdef\csname r@ftext\number\r@fcount\endcsname%
    {\message{Reference #1 to be supplied.}\writer@f#1>>#1 to be supplied.\par
     }\fi\csname r@fnum#1\endcsname}
\def\ifuncit@d#1{\expandafter\ifx\csname r@fnum#1\endcsname\relax%
\global\advance\r@fcount by1%
\expandafter\xdef\csname r@fnum#1\endcsname{\number\r@fcount}}
\let\r@fis=\refis   \def\refis#1#2#3\par{\ifuncit@d{#1}%
    \w@rnwrite{Reference #1=\number\r@fcount\space is not cited up to now.}\fi%
  \expandafter\gdef\csname r@ftext\csname r@fnum#1\endcsname\endcsname%
  {\writer@f#1>>#2#3\par}}
\def\r@ferr{\endreferences\errmessage{I was expecting to see
\noexpand\endreferences before now;  I have inserted it here.}}
\let\r@ferences=\references
\def\references{\r@ferences\def\endmode{\r@ferr\par\endgroup}}
\let\endr@ferences=\endreferences
\def\endreferences{\r@fcurr=0{\loop\ifnum\r@fcurr<\r@fcount
    \advance\r@fcurr by 1\relax\expandafter\r@fis\expandafter{\number\r@fcurr}%
    \csname r@ftext\number\r@fcurr\endcsname%
  \repeat}\gdef\r@ferr{}\endr@ferences}
\let\r@fend=\endpaper\gdef\endpaper{\ifr@ffile
\immediate\write16{Cross References written on []\jobname.REF.}\fi\r@fend}
\catcode`@=12
\citeall\refto\citeall\ref\citeall\Ref
\catcode`@=11
\newcount\tagnumber\tagnumber=0
\immediate\newwrite\eqnfile\newif\if@qnfile\@qnfilefalse
\def\write@qn#1{}\def\writenew@qn#1{}
\def\w@rnwrite#1{\write@qn{#1}\message{#1}}
\def\@rrwrite#1{\write@qn{#1}\errmessage{#1}}
\def\taghead#1{\gdef\t@ghead{#1}\global\tagnumber=0}
\def\t@ghead{}\expandafter\def\csname @qnnum-3\endcsname
  {{\t@ghead\advance\tagnumber by -3\relax\number\tagnumber}}
\expandafter\def\csname @qnnum-2\endcsname
  {{\t@ghead\advance\tagnumber by -2\relax\number\tagnumber}}
\expandafter\def\csname @qnnum-1\endcsname
  {{\t@ghead\advance\tagnumber by -1\relax\number\tagnumber}}
\expandafter\def\csname @qnnum0\endcsname
  {\t@ghead\number\tagnumber}
\expandafter\def\csname @qnnum+1\endcsname
  {{\t@ghead\advance\tagnumber by 1\relax\number\tagnumber}}
\expandafter\def\csname @qnnum+2\endcsname
  {{\t@ghead\advance\tagnumber by 2\relax\number\tagnumber}}
\expandafter\def\csname @qnnum+3\endcsname
  {{\t@ghead\advance\tagnumber by 3\relax\number\tagnumber}}
\def\equationfile{\@qnfiletrue\immediate\openout\eqnfile=\jobname.eqn%
  \def\write@qn##1{\if@qnfile\immediate\write\eqnfile{##1}\fi}
  \def\writenew@qn##1{\if@qnfile\immediate\write\eqnfile
    {\noexpand\tag{##1} = (\t@ghead\number\tagnumber)}\fi}}
\def\callall#1{\xdef#1##1{#1{\noexpand\call{##1}}}}
\def\call#1{\each@rg\callr@nge{#1}}
\def\each@rg#1#2{{\let\thecsname=#1\expandafter\first@rg#2,\end,}}
\def\first@rg#1,{\thecsname{#1}\apply@rg}
\def\apply@rg#1,{\ifx\end#1\let\next=\relax%
\else,\thecsname{#1}\let\next=\apply@rg\fi\next}
\def\callr@nge#1{\calldor@nge#1-\end-}\def\callr@ngeat#1\end-{#1}
\def\calldor@nge#1-#2-{\ifx\end#2\@qneatspace#1 %
  \else\calll@@p{#1}{#2}\callr@ngeat\fi}
\def\calll@@p#1#2{\ifnum#1>#2{\@rrwrite{Equation range #1-#2\space is bad.}
\errhelp{If you call a series of equations by the notation M-N, then M and
N must be integers, and N must be greater than or equal to M.}}\else%
{\count0=#1\count1=#2\advance\count1 by1\relax\expandafter\@qncall\the\count0,%
  \loop\advance\count0 by1\relax%
    \ifnum\count0<\count1,\expandafter\@qncall\the\count0,  \repeat}\fi}
\def\@qneatspace#1#2 {\@qncall#1#2,}
\def\@qncall#1,{\ifunc@lled{#1}{\def\next{#1}\ifx\next\empty\else
  \w@rnwrite{Equation number \noexpand\(>>#1<<) has not been defined yet.}
  >>#1<<\fi}\else\csname @qnnum#1\endcsname\fi}
\let\eqnono=\eqno\def\eqno(#1){\tag#1}\def\tag#1$${\eqnono(\displayt@g#1 )$$}
\def\aligntag#1\endaligntag  $${\gdef\tag##1\\{&(##1 )\cr}\eqalignno{#1\\}$$
  \gdef\tag##1$${\eqnono(\displayt@g##1 )$$}}
\def\eqalignno#1{\displ@y \tabskip\centering
  \halign to\displaywidth{\hfil$\displaystyle{##}$\tabskip\z@skip
    &$\displaystyle{{}##}$\hfil\tabskip\centering
    &\llap{$\displayt@gpar##$}\tabskip\z@skip\crcr
    #1\crcr}}
\def\displayt@gpar(#1){(\displayt@g#1 )}
\def\displayt@g#1 {\rm\ifunc@lled{#1}\global\advance\tagnumber by1
        {\def\next{#1}\ifx\next\empty\else\expandafter
        \xdef\csname @qnnum#1\endcsname{\t@ghead\number\tagnumber}\fi}%
  \writenew@qn{#1}\t@ghead\number\tagnumber\else
        {\edef\next{\t@ghead\number\tagnumber}%
        \expandafter\ifx\csname @qnnum#1\endcsname\next\else
        \w@rnwrite{Equation \noexpand\tag{#1} is a duplicate number.}\fi}%
  \csname @qnnum#1\endcsname\fi}
\def\ifunc@lled#1{\expandafter\ifx\csname @qnnum#1\endcsname\relax}
\let\@qnend=\end\gdef\end{\if@qnfile
\immediate\write16{Equation numbers written on []\jobname.EQN.}\fi\@qnend}
\catcode`@=12

\magnification 1200
\oneandahalfspace

\def\QB{\overline Q}
\def\PB{\overline P}
\def\@{\epsilon}
\def\itt{\int_{\tau_1}^{\tau_2}}
\def\pf{\pi_\phi}
\def\po{\pi_\Omega}
\def\pu{\pi_u}
\def\pv{\pi_v}

\line{\hfill hep-th/9512146}
\medskip

\title{CANONICAL GAUGES IN THE PATH INTEGRAL FOR PARAMETRIZED SYSTEMS}
\author{Rafael Ferraro$^{1,2}$ and Claudio Simeone$^1$}
\affil{$^1$\fceyn}
\affil{$^2$\iafedir}
\abstract

It is  well  known  that  --differing from ordinary gauge systems-- canonical
gauges are not admissible in the path integral for parametrized systems.
This is the case for the relativistic particle and  gravitation.
However, a time dependent canonical transformation can turn a parametrized
system into an ordinary gauge system. It is shown how to build
a canonical transformation such that
the  fixation  of  the  new coordinates is equivalent to the fixation of  the
original ones; this aim can be achieved only if the Hamiltonian constraint
allows  for an intrinsic global time.  Thus the resulting action, describing an
ordinary gauge  system  and  allowing  for  canonical  gauges, can be used
in the path integral  for  the quantum propagator associated with the
original variables.

\endtitlepage

\bigskip
{\bf \ 1. Introduction}
\taghead{1.}
\medskip
When the transition amplitude  for  a gauge system is written as the sum
over all histories of the exponencial of the gauge-invariant action, the path
integral diverges because of the integration over the non physical  degrees of
freedom. This difficulty can be solved by imposing gauge conditions which
select  one path from each class of physically equivalent paths.
Admissible gauges are those which can be reached from any path  by performing
gauge  transformations leaving invariant  the  action.    A gauge
transformation is generated  by the first class constraints $G_a(q^i,p_i)$,
$$[G_a,G_b]={C_{ab}}^c(q^i,p_i) G_c.\eqno(e)$$
Under a gauge transformation the action changes by an endpoint term
$$\delta_{\epsilon } {\cal S}=\left[ \epsilon^a (\tau )\left({p_i{\partial
G_a\over\partial p_i}-G_a}\right)\right] _{\tau_1} ^{\tau_2}. \eqno(g)$$
If  the constraints  are  linear  and  homogeneous  in  the  momenta,  as  in
Yang-Mills theories, the endpoint term \(g) vanishes.
If not, as it happens  with  generally  covariant systems as the relativistic
particle  and    gravitation,    the   action  is  invariant  under  a  gauge
transformation only  if  the  gauge
parameters $\@^a(\tau)$ vanish at the endpoints. Thus the
admissible gauges in the path integral  for  generally  covariant  systems are
those which can be reached from any  path  by means of a gauge transformation
mapping the boundaries onto themselves.   This is a serious limitation to the
gauge conditions to be used in  the  path  integral: canonical gauge
conditions (ie, those of the type $\chi(q,p,\tau)$  $= 0$) cannot be used with
generally covariant systems because, since the action has  not  gauge freedom
at  the boundaries, they would imply a restriction on  the  initial  and  final
quantum states.

This  difference  was taken as the main distinction  between  ordinary  gauge
systems and gene\-rally covariant systems \refto{T}.  However,
the  gauge  freedom  at  the endpoints can be recovered  if  the  action  is
modified by appropriate endpoint terms, as has been recently shown for simple
systems \refto{HTV,HT}.

We will  develop  a  method  to obtain these terms for a generic parametrized
system having  only one constraint $H$, by taking
them  as  the  consequence  of a time-dependent canonical transformation such
that one  of  the  new  momenta,  say  $P_0$,  coincides with the Hamiltonian
constraint $H$.   Then,  the  new  variables $(Q^\mu , P_\mu),$ $\mu\not= 0,$
will be observables (although not conserved), while $Q^0$ will be pure gauge.
In the new variables the
constraint is linear and homogeneous in the momenta;  thus, the action $S(Q^i
,P_i )$ will have gauge freedom  at  the  endpoints  and  a  canonical  gauge
condition will be  admissible.  In addition the canonical transformation will
be built in such  a  way  that the quantum state $|q^i>$ is equal to $|Q^i>$.
Then the action $S(Q^i,P_i)$, which is stationary on the classical trajectory
when the $Q$'s are fixed at the boundaries, will result appropriate for
computing the propagator $<q'|q>$.
\vskip2cm

{\bf \ 2. Parametrized systems}
\taghead{2.}
\medskip
The action of a parametrized system reads
$${\cal S}[q^i,p_i,N]=\int_{\tau_1}^{\tau_2}\left( p_i{dq^i\over d\tau}
- NH\right) d\tau,\eqno(31)$$
where $H$ is the null Hamiltonian and the lapse function  $N(\tau)$  is the
Lagrange multiplier enforcing the constraint $H=0$.   The  constraint implies
the existence of non-physical variables, which  leads  to  an action
with some kind of invariance or symmetry.

Under arbitrary changes of  $q$, $p$ and $N$ it is obtained
$$\delta {\cal S}=p_i\delta q^i\vert_{\tau_1} ^{\tau_2} +
\int _{\tau_1}^{\tau_2}{\left[ \left({\dot q}^i-N{{\partial H}\over
{\partial p_i}}\right)\delta  p_i-\left( {\dot  p}_i+    N{{\partial  H}\over
{\partial q^i}}\right)\delta  q^i-H\ \delta N\right] } d\tau . \eqno(32)$$
The action is stationary on the  classical  path  when  the
endpoint values of $q^i$ are fixed.

The action \(31) has two different types of invariance:

\parindent0pt 1) Invariance  under a reparametrization
$$\delta q^i= \epsilon (\tau )\ {dq^i\over d\tau },\eqno(35a)$$
\medskip
$$\delta p_i= \epsilon (\tau )\ {dp_i\over d\tau },\eqno(35b)$$
\medskip
$$\delta N={d(N \epsilon )\over d\tau }\eqno(35)$$
with $\epsilon (\tau_1)=0=\epsilon (\tau_2).$ This transformation is called a
reparametrization because it is equivalent to change $\tau$ by $\tau +\epsilon
(\tau )$ on the path given by
$q^i (\tau )$ and $p_i (\tau )$, the  integral
$$\itt N(\tau )\ d\tau $$
remaining  unchanged.        The  invariance of the   action    \(31)    under
a reparametrization  means that $\tau$ is not the time but a
physically irrelevant parameter.   When  a  system  is described by an action
like  that  of \(31), the  solutions  of  the  dynamical  equations  are  not
parametrized by $\tau$ but are
$$q^i=q^i\left( \int^{\tau}N d\tau\right)\ \ \ \ \
p_i=p_i\left( \int^{\tau}N d\tau \right) .\eqno(351)$$
So the ``proper time" $\int^{\tau}N d\tau$, instead of $\tau$, plays the role
of time.  When the equations  \(351) can be globally solved for $\int^{\tau}N
d\tau$ (ie $\int^{\tau}N d\tau = t(q,p)$), it is said that the system has a
global phase time $t(q,p)$ \refto{hac}.

\parindent0pt 2) Invariance under a gauge transformation
$$\delta_{\epsilon} q^i=  \epsilon (\tau )\ [q^i,H]\ =\ \epsilon (\tau)\
{\partial H\over \partial p_i},\eqno(341)$$
\medskip
$$\delta_{\epsilon} p_i\ =\ \epsilon (\tau )\ [p_i,H]\ =\ -\epsilon (\tau )\
{\partial H\over\partial q^i}. \eqno(342)$$
\smallskip
Then
$$\delta_\@ {\cal S}=p_i\ \delta_\@ q^i\vert_{\tau_1} ^{\tau_2} -
\int _{\tau_1}^{\tau_2}{\left[ \@(\tau )\left({\dot q}^i{{\partial H}\over
{\partial q^i}} + {\dot  p}_i{{\partial  H}\over
{\partial p_i}}\right) +H\ \delta_\@ N\right] } d\tau . \eqno(732)$$
\medskip
As $\delta_{\epsilon } N$ cannot be generated by $H$, it can be defined
$$\delta_{\epsilon } N=\dot\epsilon ,\eqno(736)$$
and then
$$\delta_\@   {\cal  S}\ =\ p_i\ \delta_\@   q^i\vert_{\tau_1}  ^{\tau_2}
-\ \int_{\tau_1}^{\tau_2}{d\over d\tau} (\@ H) d\tau \
=\ \left[    \epsilon (\tau )\left({p_i{\partial  H\over\partial p_i}-H}
\right)\right] _{\tau_1} ^{\tau_2}. \eqno(737)$$
On the classical path, where Hamilton equations hold,
the reparametrization \(35a)-\(35) is  equivalent  to  a  gauge transformation
with parameter $N\@$  and  the boundary restrictions $\epsilon (\tau_1)=0=
\epsilon(\tau_2)$.

If the constraint  $H$  is  not  linear and homogeneous in the momenta, as is
usual  when  one  deals    with  parametrized  systems,  the  action  is  not
gauge-invariant unless the restrictions
$$\epsilon (\tau_1)=0=\epsilon (\tau_2)\eqno(2@)$$
are added.

Gauge invariance is usually  regarded  as the consequence of the existence of
spurious  degrees of freedom.   However,  gauge  invariance  of  parametrized
systems is related to reparametrization invariance;    ie    the physically
irrelevant variable is not a canonical variable but is the parameter $\tau .$
$\tau$  is  not the time but the time  can  be  hidden  among  the  dynamical
variables.  This is the case when a global  phase  time  exists  (the  Jacobi
action is an example of a parametrized action which has  not  a  global phase
time \refto{LBY}).  As a result, the path integral for such a system does not
depend on $\tau_1 ,\tau_2 ,$ but only  on  the  initial  and  final values of
$q^i$. Hence,  the path integral for a parametrized system corresponds
to the probability $<{q^i}'\vert q^i>.$

The restrictions \(2@) make impossible to fix the gauge in the path integral
by imposing conditions
on the canonical variables of the form
$$ \chi(q,p,\tau )=0,\eqno(chi)$$
(``canonical gauges").  This type of gauge conditions are not admissible when
the  constraints  are  not  linear  and    homogeneous  because,  due  to  the
restrictions \(2@), there is no gauge freedom at the endpoints, and then \(chi)
would imply a restriction on the initial and final quantum states \refto{T}.

Admissible gauges in the path integral are those which can be carried  to
$\chi =0$ by means of a gauge  transformation leaving the action invariant.
Let us consider a trajectory which differs from the condition $\chi=0$ by an
infinitesimal quantity $\Delta$; the gauge transformation
which makes the variables reach the gauge condition must be such that
$$\delta_\@ \chi=-\Delta .\eqno(738)$$
In order to have  only  one solution $\@ (\tau)$ with the boundary conditions
\(2@), \(738) should be  a  second  order
differential  equation  in  the parameter $\@$.  Since  $\delta_\@
N=\dot\@ ,$ the most obvious gauge condition could be given by a  function of
$\dot N$, namely \refto{T}
$$\chi=\dot N =0.\eqno(40)$$
Any particular choice of $N(\tau )$ can be carried
to $\dot N=0$ by successive infinitesimal
gauge transformations  $  \delta_\@ N=\dot\@ ;$ these transformations are
possible because there are no restrictions on $\dot\@  ,$  but only on
$\@$ at the endpoints. Gauges like \(40) are called ``derivative gauges".
Although the gauge condition \(40) does not fix  the  value  of $N$, but only
says that $N$ is constant on the trajectory, the  value  of $N$ is determined
by  the variational principle itself when the data at $\tau_1$  and  $\tau_2$
are enough for knowing the global phase time $t(q,p)$ at the  endpoints.    In
fact, $N = \Delta t /\Delta\tau$.  So no ambiguities are left on the classical
trajectory.
\bigskip
The practical  value  of  having  linear and homogeneous constraints led to
distinguish these ordinary gauges systems from
all others by calling them systems with {\it internal gauge
symmetries}.  However, there  is not a true conceptual difference between both
classes  of  systems:   internal  gauge  symmetry  can  be  no  more  than  a
consequence  of  a  particular  choice  of   variables,  and  an  appropriate
transformation    $(q^i,p_i)    \rightarrow   (Q^i,P_i)$  can  eliminate  the
restrictions   $\epsilon  (\tau_1)=0=\epsilon  (\tau_2)$  on  the  admissible
gauges, allowing us to impose canonical gauge conditions in the path integral
\refto{HT}.
\vskip2cm
\vfill
\eject
{\bf  3. The endpoint terms}
\medskip
\taghead{3.}

As we have seen, the  general  form  of  the  variation of the action under a
gauge transformation is that of an  endpoint  term  (see \(737)).    It  is
then
possible to achieve gauge freedom at the  endpoints  by  means  of  including
appropriate endpoint terms in the action.  These terms have been obtained in
\Ref{HTV} for the  parametrized free particle and the free relativistic
particle.

In this work we develop a method that gives the appropriate endpoint terms for
a para\-metrized system in a general way, by seeing them as  a  consequence  of
having performed a suitable canonical transformation.  If the endpoint terms
are called
$B$, the gauge-invariant action of a parametrized system reads
$$S[q^i,p_i,N]=\int _{\tau_1}^{\tau_2}\left( p_i{dq^i\over d\tau}-NH\right) d
\tau +B ,\eqno(401)$$
where, as it follows from \(737), it is clear that it must be
$$\delta_{\epsilon}  B=   \left[ - \epsilon (\tau ) \left({p_i{\partial  H\over
\partial p_i}-H}\right)\right]_{\tau_1} ^{\tau_2} \eqno(41)$$
to have $\delta_{\@} S=0$ for any gauge transformation.
\bigskip
Let us consider a complete solution $W(q^i,\alpha_\mu,E)$ of the
$\tau$-independent Hamilton-Jacobi equation
$$H\left( q^i,{\partial  W\over\partial q^i}\right) =E.\eqno(52)$$
If $E$ and the integration constants  $\alpha_\mu$  are  matched  to
the  new  momenta
$\PB_0$ and $\PB_\mu$  respectively, then $W(q^i,\overline P_j)$ can be
regarded  as    the    generator   function  of  a  canonical
transformation $(q^i,p_i) \rightarrow (\QB^i,\PB_i)$ defined by the equations
$$p_i = {\partial W\over\partial q^i}$$
$$\overline Q^i = {\partial W\over\partial\overline P^i}$$
$$\overline K = N\overline P_0. \eqno(54)$$
As the transformation is canonical,  it is clear that
$$[\overline Q^\mu ,\overline P_0] = [\overline Q^\mu , H]=0$$
$$[\overline    P_\mu ,\overline    P_0] = [\overline   P_\mu ,    H] = 0,
\eqno(551)$$
which means that  $\overline  Q^\mu$  and  $\overline  P_\mu$  are
(conserved) observables describing the {\it reduced} system.

The dynamical evolution for $\QB^0$
$${d\QB^0\over d\tau}=[\QB^0,\overline K]=N[\overline Q^0,\overline P_0]=N$$
is solved by $\QB^0=\int^{\tau}Nd\tau .$ If $\QB^0$ is globally well defined,
then $\QB^0$ is a global phase time.

The action
$$\overline  S\ [\overline Q^i,\overline P_i,  N]=  \itt \left(\overline P_i
{d\overline Q^i\over d\tau}-N\overline P_0\right) d\tau \eqno(56)$$
describes    a   parametrized  system  with  a  constraint  which  is  linear
and homogeneous  in the momenta.  Therefore the action $\overline S$ has gauge
freedom at the boundaries,  and  does  not  need  endpoint  terms.  Canonical
gauges are then admissible in a path integral with the action $\overline S$.
A canonical gauge  can  be  chosen  to  be  $\chi = \overline Q^0 - g(\tau)$,
meaning that $N(\tau)=g'(\tau)$ on the classical trajectory.

The action $\overline S$ can be related with ${\cal S}$ by noting that
$$p_i dq^i =\ d(W-\QB^i \PB_i)+\PB_i d\QB^i,$$
as it follows from \(54). Then
$$\overline  S
=\itt \left( p_i {dq^i\over d\tau}-NH\right) d\tau +\left[\QB^i
(q^i,p_i)\PB_i(q^i,p_i)-W\right]_{\tau_1} ^{\tau_2} \eqno(63)$$
and the endpoint terms making the action ${\cal S}$ gauge-invariant are
$$\overline B=\left[\QB^i(q^i,p_i)\PB_i(q^i,p_i)-W\right]_{\tau_1}  ^{\tau_2}
;\eqno(64)$$
the property \(41) is straightforwardly verified by these terms.
\vskip2cm

{\bf \ 4. The variables to be fixed at the endpoints}
\medskip
\taghead{4.}
We  have  succeeded in  identifying the reduced  system,  described  by the
coordinates and  momenta  $(\QB^\mu,  \PB_\mu  )$,  and in getting the action
\(56)  which  has  gauge  freedom  at  the  boundaries.
The added endpoint  terms do not change the dynamical equations, but  change
the
quantities to  be  fixed  at  the endpoints in order to get the trajectories
from the variational principle.
The action \(56) requires fixing the $Q^i$'s   at the
endpoints  (actually  only the  $Q^\mu$'s should be fixed, since $\PB_0  =  0$
on  the  classical trajectory). So $\overline  S$  is appropriate to
compute  $<{\QB^i}'\vert\QB^i> = <{\QB^\mu}',{\QB^0}'\vert\QB^\mu,\QB^0>$
-remind that $\QB^0$ is the global phase time- by means of a path integral
allowing  for  canonical  gauges.  However, our aim
is to compute $<{q^i}'\vert q^i>$, instead of $<{\QB^i}'\vert\QB^i>$;
but $<{\QB^i}'\vert\QB^i>$ is  not  equal to $<{q^i}'\vert
q^i>,$ because the choice of the $\QB^i$'s  does  not fix the same
quantum state that the choice of the $q^i$'s does. In fact,
equations \(54) and \(551) tell us that the variables $(\QB^\mu ,\PB_\mu )$ are
conserved on  classical  trajectories.  While any classical trajectory can be
characterized  by
the choice of the $q^i$'s at the endpoints, in the new variables this is done
by
the choice  of  the  conserved  observables  $(\QB^\mu  ,\PB_\mu  ),$ and the
$\QB^0$'s at the  endpoints.   Thus, new and original variables play different
roles in characterizing states  or  histories, and the amplitudes $<{q^i}'\vert
q^i>$ and $<{\QB^i}'\vert\QB^i>$ have different meanings.

Still, one can look for a propagator equal to $<{q^i}'\vert q^i>$ by performing
a canonical transformation $(\QB^\mu,\PB_\mu) \rightarrow (Q^\mu  ,P_\mu)$
in the reduced space.  If  this  transformation is $\tau$-dependent
the Hamiltonian will change; then the observables $Q^\mu$ will not be
conserved, and  one  could succeed in getting the wished propagator.
Let us consider the canonical transformation  generated by
$$F(\QB, P)\ =\ P_0\ \QB^0+f(\QB^\mu ,P_\mu ,\tau ). \eqno(70)$$
Then
$$H=\PB_0={\partial F\over\partial\QB^0}=P_0$$
$$\QB^0={\partial F\over\partial\PB_0}=Q^0 .\eqno(701)$$
The transformation  $(\QB^\mu
,\PB_\mu) \rightarrow (Q^\mu  ,P_\mu)$ is generated by $f(\QB^\mu ,P_\mu ,
\tau ).$
$Q^\mu$ and $P_\mu$ are observables, because their Poisson  brackets  with
$P_0=H$ are zero, but are  not  conserved because their evolution is governed
by the non-zero Hamiltonian
$$K=\overline           K+{\partial        F\over\partial\tau}=NP_0+{\partial
f\over\partial\tau}\eqno(702)$$
($h\equiv\partial f/\partial\tau$ is the Hamiltonian for the reduced system).

The additional endpoint term
$$\left[        Q^iP_i-F\right]_{\tau_1}^{\tau_2}=\left[                Q^\mu
P_\mu-f(\QB^\mu ,\PB_\mu ,\tau )\right]_{\tau_1}^{\tau_2}$$
depends only on  observables;    then it is gauge-invariant.  This means that
the action
$$S[Q^i,P_i,N] = \int \left( P_i{dQ^i\over d\tau}\ -\ N\ P_0\ -\ {\partial
f\over\partial\tau} \right) d\tau  \eqno(811)$$
also  has  gauge freedom at  the  endpoints  (which  appears to be obvious  if
we regard  that after the new canonical  transformation  generated  by  $f$
the
constraint remains linear in the momenta).

The  action $S[Q^i,P_i,N]$ describes a non parametrized system  with  internal
gauge symmetry.  For this system $\tau$ {\it  is} the time, but $Q^0$ is pure
gauge (of course,  the  roles  of  $\tau$ and $Q^0$ are interchangeable since
$Q^0$ is a global phase time).  Then, the gauge can be fixed in  the  path
integral  by  means  of a canonical gauge.

The action \(811) is appropriate to compute the amplitude $<{Q^i}',\tau_2\vert
Q^i,\tau_1 >$
$$<{Q^i}',\tau_2\vert Q^i ,\tau_1>\ =\
\int {DQ^0  DP_0  DQ^\mu  DP_\mu DN \delta (\chi ) {\vert
[\chi ,P_0] \vert} e^{iS} }, \eqno(84)$$
where $\chi$ is any admissible canonical gauge, and ${\vert
[\chi ,P_0] \vert}$ is the Fadeev-Popov determinant. Let us pay attention to
the fact that  this
amplitude  depends  on  $\tau_1$  and $\tau_2$, because the new action $S$ is
gauge-invariant but is  not  invariant  under reparametrizations.  Of course,
the path integration in  eq.\(84)  is  nothing  but the path integral for the
reduced system:   the  functional integration on $N$ enforces the path to lie
on the constraint hypersurface,
then by using $\chi\equiv  Q^0-g(Q^\mu  ,\tau )$ --which gives the
endpoint values of $Q^0$ in terms of the endpoint values of
$Q^\mu$ and $\tau$--, one integrates in
$Q^0$ and $P_0$ to  obtain  $<{Q^\mu}',  \tau_2\vert  Q^\mu  ,\tau_1>$.

In the propagator \(84), $S$ is related to the original action by
$$S=\itt \left(  p_i{dq^i\over d\tau  }-NH\right)  d\tau + B  , \eqno(79)$$
where
$$B\equiv    \left[   \QB^i  \PB_i    -W+Q^\mu    P_\mu    -f\right]_{\tau_1}
^{\tau_2}\eqno(790)$$
can be expressed as a function of the original canonical variables $q^i$
and $p_i$.

As the generator $f(\QB^\mu ,P_\mu,\tau)$ has not been defined yet, one can
try to define it in  such  a  way  that  $<{Q^i}',\tau_2\vert  Q^i,\tau_1 >$
($= <{Q^\mu}',\tau_2\vert Q^\mu,\tau_1  >$) coincides with
$<{q^i}'\vert q^i>$. In order to reach this aim one must check that
\smallskip
\parindent0pt {\bf 1.} The constraint is  such  that  it  admits  a canonical
gauge $\tilde\chi$  (satisfying $[\tilde\chi,H]\not\approx 0$ \refto{dirac})
depending only  on  $\tau,  q^i$.
Then $\tilde \chi =0$ defines $\tau$ as a function  $\tau = \tau(q^i)$.
If so, one  says  that  there exists an  {\it  intrinsic   time} \refto{K2}.
\medskip
If this requirement is fulfilled, one chooses the generator $f(\QB^\mu,P_\mu)$
in such a way that
\smallskip
\parindent0pt {\bf 2.}  The  gauge-invariant  coordinates $Q^\mu$  behave as
coordinates on the surface $\tilde \chi=0$, so meaning that a particular choice
of  $Q^\mu$ and  $\tau$  defines  a point $q^i$ in the original configuration
space.

In that case, $|Q^\mu \tau> =  |q^i>$.    However, the path integral \(84) is
gauge  invariant;   then, not only $\tilde  \chi$  but  any  canonical  gauge
condition can be used in \(84). So the path integral \(84) is equal to the
propagator  $<{q^i}'|q^i>$  when  the  generator  $f(\QB^\mu,P_\mu)$  is chosen
according  to  the prescription 2.

A practical way  to  understand  the prescription 2 comes of considering
the eq.  \(79).  In fact, while the  action  $S$ is stationary on the
classical  trajectory  when  the values of $Q^\mu$ are fixed at $\tau_1$  and
$\tau_2$, the action ${\cal S}$ on the right hand side requires the  fixation
of  the $q^i$'s.  In order that both set of variables are equivalent  in  the
gauge $\tilde \chi=0$, the  generator  $f(\QB^\mu,P_\mu)$ should be such that
the endpoint terms  vanish  on  the  constraint  hypersurface when the gauge
$\tilde \chi=0$ is used,

$$B\equiv    \left[   \QB^i  \PB_i    -W+Q^\mu    P_\mu    -f\right]_{\tau_1}
^{\tau_2}\ {\vert_{P_0=0,\tilde \chi=0}}\ =\ 0.\eqno(7901)$$

If so, the paths will be weighted by the original action ${\cal S}$.

In the next Section we  shall  apply this way of choosing the generator
$f$ in several examples.

\vskip2cm
{\bf 5. Examples}
\bigskip
{\bf 5.a  Parametrized free particle}
\smallskip
\taghead{5. }
This is the system obtained when the time $t$ is included among the dynamical
variables of a free particle.    The parametrized particle is then  described
by  the  original variables $q$  and  $p$,  plus  $t$  and  its  conjugate
momentum $p_t$.  The action of this system is
$${\cal S}\left(  q,p,t,p_t  ,N\right)  =\int  \left(  p\dot  q+p_t\dot  t -
NH\right) d\tau \eqno(128)$$
with
$$H=p_t+{p^2\over 2m}.\eqno(130)$$
(By solving the  constraint  for  $p_t$  the  action  of  a  non relativistic
particle  is  recovered.)  A  complete  solution  of  the  $\tau$-independent
Hamilton-Jacobi equation is
$$W(q,t,\PB ,\PB_0)=\PB q+\left( \PB_0-{\PB ^2\over 2m}\right) t.\eqno(131)$$
The gauge defining an intrinsic time is $\tilde\chi\equiv t-T(\tau)$ (for any
monotonous function $T$)
and the  appropriate  function  $f$  making the endpoint terms vanish in this
gauge is
$$f(\QB ,P,\tau )=\QB P+{P^2\over 2m}T(\tau ).$$
The    original  variables  $(q^i,p_i)$  are then  related  to  the  new ones
by
$$\eqalign{Q^0 &=t\cr
Q &=q-{P\over m}(t-T(\tau))\cr
p_t &=P_0-{P^2\over 2m}\cr
p &=P.\cr} \eqno(133)$$
On the constraint surface  $P_0=0$ the endpoint terms read
$$B = \left[ -{p^2\over 2m}(t-T(\tau))\right]_{\tau_1}^{\tau_2} \eqno(134)$$
and vanish in the gauge  $\tilde\chi
=0$. The amplitude  $<q'\ t'\vert q\ t>$ can be written as
$$\int Dt Dp_t  DqDpDN\delta  (\chi)\vert  [\chi  ,H]\vert
\ \ e^{i\itt\left(  p_t{dt\over  d\tau}+p{dq\over  d\tau}-NH\right)    d\tau
-
i\left[ {p^2\over 2m}(t-T(\tau))\right]_{\tau_1} ^{\tau_2}}\eqno(135)$$
but, since the action is gauge-invariant, the amplitude can be  computed  in
{\it any} canonical  gauge.    For instance, one can choose $\chi\equiv t=0,$
and obtain
$$\eqalign{    <q'\ t\vert  q\ t>&=\int  DqDp \ \ e^{i\itt  \left[    p{d\over
d\tau}(q+{p\over m}T(\tau)) -{p^2\over\ 2 m}{dT\over d\tau}\right]    d\tau}\cr
&=\int  DQ DP\ \  e^{i\itt  \left(  P{dQ\over  d\tau  }   -{P^2\over  2m}{dT
\over d\tau}\right)   d\tau} \cr}  \eqno(137)$$
The endpoint values of $Q$ and $\tau$ are related to the  endpoint  values of
$q$  and $t$ by the gauge condition $\tilde\chi\equiv t-T(\tau )=0,$ in which
the endpoint terms vanish:
$$Q\vert_{\tilde\chi=0} =q,$$
$$T(\tau )\vert_{\tilde\chi=0} =t.\eqno(138)$$
The  path  integral  for  the  free  particle  is then recovered, as could be
expected from the fact that the reduced system has the Hamiltonian
$$h={P^2\over 2m}.$$
Cf \Ref{HTV}.

\vskip.7cm
{\bf 5.2 Relativistic free particle}
\smallskip
The  Hamiltonian constraint of this system is
$$H={1\over 2m}({p_0}^2-p^2-m^2).\eqno(1390)$$
The $\tau$-independent Hamilton-Jacobi equation has two different solutions:
$$W_\pm                                (x,x^0,\PB,\PB_0)                 =\PB
x\pm x^0{\sqrt{\PB^2+2m{\PB_0}+m^2}}.\eqno(1400)$$
The gauge  $\tilde\chi\equiv  x^0-T(\tau)$  defines  an  intrinsic  time.   The
generator $f$ making  the  endpoint  terms  vanish  when  $\tilde\chi =0$ and
$P_0=0$ is
$$f(\QB ,P,\tau )=\QB P\mp T(\tau )\sqrt{P^2+m^2} .\eqno(1139)$$
The relation between original and new variables is
$$\eqalign{Q^0 &=\pm{mx^0\over{\sqrt{P^2+2m{P_0}+m^2}}}\cr\cr
Q &=x\pm
{Px^0\over{\sqrt{P^2+2m{P_0}+m^2}}}\mp{PT(\tau )\over\sqrt{P^2+m^2}}\cr\cr
p_0 &=\pm {\sqrt{P^2+2m{P_0}+m^2}}\cr\cr
p &=P.\cr\cr}. \eqno(139)$$
On the constraint surface  the endpoint terms are
$$B = \left[ \mp{m^2(x^0-T(\tau))\over\sqrt{P^2+m^2}}\right]_{\tau_1}
^{\tau_2},\eqno(140)$$
and vanish in the gauge $\tilde\chi=0.$
The amplitude $<x'\ {x^0}'\vert x\ x^0>$ is equal to
$$\int Dx^0\ Dp_0\  Dx\ Dp\ DN\ \delta  (\chi  )\ \vert  [\chi
,H]\vert\ \ e^{i\itt\left( p_0{dx^0\over d\tau} + p{dx\over d\tau} - N H\right)
d\tau\mp  im^2\left[  {x^0-  T(\tau  )\over  \sqrt  {p^2+m^2}}\right]_{\tau_1}
^{\tau_2}} ,\eqno(141)$$
but  it  can  be  computed  in any canonical gauge; by choosing $\chi\equiv
x^0=0$ the following result is obtained:
$$<x'\ {x^0}'\vert x\ x^0>=\int  DxDp\ \ e^{i\itt p{dx\over  d\tau}
d\tau\pm  i\left[ m^2  T(\tau  )\over  \sqrt  {p^2+m^2}\right]_{\tau_1}
^{\tau_2}} ,\eqno(143)$$
where, with the choice $\chi\equiv x^0=0$,
$$\eqalign{\itt p{dx\over  d\tau}
d\tau\pm  \left[ {m^2  T(\tau  )\over  \sqrt  {p^2+m^2}}\right]_{\tau_1}
^{\tau_2}&=\itt  \left[    p{d\over   d\tau}\left(  x\mp{pT(\tau)\over  \sqrt
{p^2+m^2}}\right)\pm \sqrt  {p^2+m^2}{dT\over d\tau}\right]
d\tau \cr\cr
&=\itt \left[ P{dQ\over  d\tau}\pm  \sqrt  {P^2+m^2}{dT\over  d\tau  }\right]
d\tau .\cr\cr}\eqno(144)$$
The endpoint  values  of $Q$ and $\tau$ are related to those of $x^0$ and $x$
by the gauge condition such that the endpoint terms vanish:
$$Q\vert_{\tilde\chi=0} =x,$$
$$T(\tau )\vert_{\tilde\chi=0} =x^0.\eqno(145)$$
As it could be expected,  the Hamiltonian of the reduced system is
$$h= \mp\sqrt  {P^2+m^2},$$
and the path integral for a free relativistic particle is obtained.
\vskip.7cm

{\bf 5.3 A more general constraint}
\smallskip
A complete solution of  the  Hamilton-Jacobi  equation is mostly difficult to
obtain.  There is a  simple  case  in  two  dimensions,  generalizing the
former examples, which can be applied to some minisuperspace models in
cosmology. Let us consider a Hamiltonian constraint
$$H(\phi ,\Omega ,\pf ,\po )=g(\phi ,\Omega ) (\pf^2-\po^2) +V(\phi ,\Omega)
\eqno(167)$$
where  $g(\phi  ,\Omega  )$  and  $V(\phi  ,\Omega)$  are  positive  definite
functions.

Let us use null coordinates defined as
$$u=R(\phi+\Omega ),\ \ \ v=R(\phi-\Omega)$$
where $R$ is some monotonous function. Then
$${1\over  4}  (\pf^2-\po^2)=    R'(\phi+\Omega    )R'(\phi-\Omega)\pu    \pv
.\eqno(170)$$
In the case that the potencial can be written as
$$V(\phi        ,\Omega )=g(\phi             ,\Omega )L_1
(\phi+\Omega)L_2(\phi-\Omega)\eqno(168)$$
one can factorize out  a  positive definite factor in $H$ by choosing $R'$ as
the positive definite function $L_1$:
$$H=4g(\phi ,\Omega)L_2(\phi-\Omega   )  L_1(\phi-\Omega  )\left[
\pu\pv+{1\over 4}{L_2(\phi-\Omega   )\over  L_1(\phi-\Omega  )}\right] .
\eqno(172)$$
The function  in brackets is
a constraint $H'$ equivalent to $H$, since  the product of the three other
factors in \(172) is positive definite.  Therefore  the  canonical
transformation can
be  generated  by  means  of a complete solution  of  the  $\tau$-independent
Hamilton-Jacobi equation associated with
$$H'=\pu\pv+{1\over 4}L(v)$$
where  $L(v)\equiv{L_2(\phi-\Omega  )/ L_1(\phi-\Omega )}.$ In such a case
$P_0$ will not be $H$ but $H'.$ Anyway, the constraint $H$ will be linear and
homogeneous  in  the  new momentum $P_0$. The new variables $Q^\mu$  and
$P_\mu$  will be observables, because their Poisson brackets with $H$ will be
zero on the constraint surface.

The generator function $W$ is
$$W=\PB u+{\PB_0\over \PB}v-{1\over 4\PB}\int L(v) dv.\eqno(174)$$
The gauge  $\tilde\chi\equiv  v-T(\tau  )$  defines an intrinsic time.  The
generator function $f$ is
$$f(\QB ,P,\tau )=\QB P+{1\over 4 P}\int L(T)dT\eqno(1744)$$
and the new variables are related to $\{u,\pu ,v,\pv\}$ by
$$\eqalign{Q^0=&{v\over P}\cr
Q=& u-{P_0\over P^2}v+{1\over 4P^2}\int_{T(\tau)} ^v L(T)dT  \cr
\pu=& P\cr
\pv=& {P_0\over P}- {L(v)\over 4P}.\cr}\eqno(1745)$$
On the constraint surface $P_0=0$ the endpoint terms read
$$B=\left[       {1\over    2P}\int_{T(\tau)}^v L(T)dT\right]_{\tau_1}
^{\tau_2},\eqno(1746)$$
and clearly vanish in the gauge $\tilde\chi = 0.$
The boundaries in the path integral are given by
$$T\vert_{\tilde\chi=0}=v=R(\phi -\Omega ),$$
$$Q\vert_{\tilde\chi=0,P_0=0}=u=R(\phi +\Omega ).$$

\vskip2cm

\centerline{\bf 6. Conclusions}
\bigskip
In \Ref{HTV} it has been signaled that a generally covariant system and an
ordinary gauge system are  not  conceptually different.  In fact, differences
between both kinds of systems,  which seemed to be an obstruction to the use
of canonical gauges in the path integral for generally covariant systems
\refto{T}, can be  saved by improving the action principle  with
appropriate endpoint terms \refto{HTV,HT}.  In this way the action is endowed
with gauge freedom at the boundaries.

The improved action  can be still modified by the addition of gauge invariant
endpoint terms.  One can take advantage of this possibility
to build  the  endpoint  terms  in  such  a  way that they cancell out on the
constraint hypersurface  when  a  gauge  defining  an intrinsic time
--$\tau=\tau(q^i)$--  is used.
This means that the dynamical variables to  be fixed in the
variational principle for the improved and the original  action respectively,
define the same physical state in both cases, so warranting that the improved
action  can  be  used  in the path integral to  compute the quantum propagator
$<{q^i}'\vert q^i>$ for  the  original variables.

However,  not  all   systems  have  an  intrinsic time.   For  instance,  the
constraint of an ideal clock \refto{BF}
$$H= -p^2 + q$$
does not  admit  a  gauge  condition  $\chi  (q,\tau  )$  because $[\chi,H] =
- p\ \partial\chi/\partial  q$  vanishes  on the constraint surface
when $p=0$.   In  this  case  the solution of the $\tau$-independent Hamilton
-Jacobi equation is
$$W\ =\pm{2\over 3}\ (q - \PB_0)^{3/2},$$
and the global phase time $\QB^0$ results in
$$\QB^0={\partial W\over\partial \PB_0}=\pm\sqrt{q - \PB_0}\ =\ p.$$
Then the gauge choice necessarily involves the momentum $p$.  In this case
it is said that the system has an {\it extrinsic} time \refto{K2}.

\bigskip
The endpoint terms can be seen as the consequence of a $\tau$-dependent
canonical transformation; we  have shown how this transformation can be
generated in the case of a parametrized system with a constraint  $H$.
In this kind of systems, the constraint means that the time parameter
$\tau$  has no physical meaning.  The canonical transformation that generates
the appropriate  endpoint  terms is such that  $\tau$ is the
time in the new system, while one  of  the  dynamical  variables  --$Q^0$--
is pure gauge, as it happens in ordinary gauge systems.

Finally we have shown  that the procedure to find the generator function $W$,
leading to the identification of  the  reduced  space,  can be facilitated by
appropriately scaling the constraint.
\bigskip

\references

\bigskip

\refis{T} C. Teitelboim, Phys.Rev. D {\bf 25}, 3159 (1982).

\refis{HTV} M.   Henneaux, C.  Teitelboim and J.  D.  Vergara, Nucl.Phys. B
{\bf 387}, 391 (1992).

\refis{HT} M.  Henneaux and C.  Teitelboim, {\it Quantization of Gauge
Systems,}
Princeton University Press, New Jersey (1992).

\refis{dirac} P.    A.  M.  Dirac, {\it Lectures on Quantum Mechanics}, Belfer
Graduate School of Science, Yeshiva University, N.Y., 1964.

\refis{K2} K.   V.    Kucha\v r,  in  {\it  Proceedings  of the 4th Canadian
Conference on General Relativity  and \ \  Relativistic  Astrophysics.}  G.
Kunstatter, D. Vincent, J. Williams, World Scientific (1992).

\refis{BF} S.  C. Beluardi and R.  Ferraro, Phys.Rev. D {\bf 52}, 1963 (1995).

\refis{LBY} C.    Lanczos,  {\it  The  Variational  Principle  of Mechanics},
(Dover, New York,  1986);   J.  D.  Brown and J.  W.  York, Phys.Rev.  D {\bf
40}, 3312 (1989).

\refis{hac} P. H\'aj\'\i cek, Phys.Rev. D {\bf 34}, 1040 (1986).

\endreferences

\end